\documentclass[12pt]{article}

\usepackage{amsmath}
\usepackage{latexsym}
\newcommand\be{\begin{equation}}
\newcommand\ee{\end{equation}}
\newcommand\bea{\begin{eqnarray}}
\newcommand\eea{\end{eqnarray}}
\newcommand\beas{\begin{eqnarray*}}
\newcommand\eeas{\end{eqnarray*}}

\def\tr{{\rm Tr}}

\begin{document}
\title{
Non supersymmetric strong coupling background from the large N quantum mechanics of two matrices coupled via a Yang-Mills interaction }

\author{\\
Jo\~ao P. Rodrigues\footnote{Email: joao.rodrigues@wits.ac.za} ~and Alia Zaidi \\
\\
National Institute for Theoretical Physics \\
School of Physics and Centre for Theoretical Physics \\
University of the Witwatersrand, Johannesburg\\
Wits 2050, South Africa 
\\
}

\maketitle

\begin{abstract}
We derive the planar large $N$ non-supersymmetric background of the quantum mechanical hamiltonian of two hermitean matrices coupled via a Yang-Mills interaction, in terms of the density of eigenvalues of one of the matrices. This background satisfies an implicit non linear integral equation, with a perturbative small coupling expansion and a solvable large coupling solution, which is obtained. The energy of system and the expectation value of several correlators are obtained in this strong coupling limit. They are free of infrared divergences. 
\end{abstract}

\newpage

\noindent
\section{Introduction}

The study of multi-matrix models\footnote{By matrix models we mean integrals over matrices or the quantum mechanics of matrix valued degrees of freedom}, 
and particularly their large $N$ limit \cite{'t Hooft:1973jz}, is of great interest. It is well known, for instance, that the large $N$ limit of their description of D$0$ branes \cite{Polchinski:1995mt} has been conjectured to provide a definition of $M$ theory \cite{Banks:1996vh}. 
In the context of the AdS/CFT duality \cite{Maldacena:1997re}, \cite{Gubser:1998bc}, \cite{Witten:1998qj},  
due to supersymmetry and conformal invariance, correlators of supergravity and $1/2$ BPS states reduce to calculation of free matrix model overlaps 
\cite{Lee:1998bxa}, \cite{Corley:2001zk} or consideration of related matrix hamiltonians \cite{Berenstein:2004kk}. 
For stringy states, in the context of the BMN limit \cite{Berenstein:2002jq} and $\cal{N}$ $=4$ SYM, similar considerations apply 
\cite{Constable:2002hw}, \cite{Beisert:2002ff}, \cite{deMelloKoch:2003pv}. 
A plane-wave matrix theory \cite{Kim:2003rza} is related to the $\cal{N}$  $=4$ SYM dilatation operator  
\cite{Beisert:2004ry}. Recently, multi-matrix, multi-trace operators with diagonal free two point functions have been identified
\cite{Brown:2007xh}, \cite{Bhattacharyya:2008rb}. In earlier works \cite{Eguchi:1982nm}, \cite{Bhanot:1982sh}, \cite{Gross:1982at} 
it has been argued that $QCD$ can be reduced to a finite number of matrices with quenched momenta.

In this communication we will consider the quantum mechanics of two hermitean matrices with harmonic potentials, 
interacting through a standard Yang-Mills quartic potential. 

We will use the approach, first developed in \cite{Donos:2005vm}, of treating one of the matrices, which generates the large $N$ planar background, in a coordinate representation, and the other in a creation/anihilation basis. This was done in the context of $1/2$ BPS states and a dual free harmonic Hamiltonian 
\cite{Corley:2001zk}, \cite{Berenstein:2004kk}
with the matrix generating the background being the holomorphic component of a complex matrix. A precise phase space identification between the collective density description of the dynamics of this matrix \cite{Jevicki:1979mb}, and the droplet description of the LLM \cite{Lin:2004nb} metric is obtained \cite{Donos:2005vm}.  
The generalization of this approach to include $g_{YM}$ interactions was developed in \cite{Rodrigues:2005ec}. 
By considering the planar background generated by of one of the two Hermitean matrices, further properties of the spectrum were established in \cite{Cook:2007et}.

In  \cite{Donos:2005vm}, \cite{Rodrigues:2005ec} and \cite{Cook:2007et}, a supersymmtric setting was always assumed, allowing one to consistently neglect normal ordering terms. As a result, the planar background is harmonic, and $g_{YM}$ independent.

In this communication we explore the consequences of not requiring supersymmetry, and establish a $g_{YM}$ dependent density description of the planar 
large $N$ limit of the system. The communication is organized as follows: in Section $2$, the hamiltonian of the two interacting harmonic matrices is introduced. 
It is shown how a Bogoliubov transformation brings the hamiltonian to a form quadratic in the creation/annhiliation oscillators of one of the matrices, with frequencies having a square root dependence on the eigenvalues of the other matrix \cite{Rodrigues:2005ec} \footnote{See also \cite{Berenstein:2005jq}}. 
The sector of the Hamiltonian contributing to the planar large $N$ background is identified. In Section $3$, the planar background, in terms of the density of one 
the matrices, is obtained implicitly through a non-linear integral equation. In Section $4$, the perturbative expansion in $\lambda=g_{YM}^2$ is described, and the first order correction obtained and shown to agree with perturbation theory. In section $5$, the strong coupling limit of the background is obtained explicitly. In Section $6$, it is argued that this strong coupling limit is indeed the background for two free hermitean matrices interacting through the Yang-Mills potential. This solution is free of infrared divergences. Section $7$ is reserved for a summary and discussions.       

\noindent
\section{Model Hamiltonian and planar large $N$ sector}

We consider the quantum mechanical hamiltonian 

\be\label{Ham12}
\hat{H}
\equiv
\frac{1}{2} Tr (P_1^2 ) + \frac{w^2}{2}  Tr (X_1^2) + \frac{1}{2} Tr (P_2^2 )+ \frac{w^2}{2}  Tr (X_2^2)
- g_{YM}^2 \tr ([X_1,X_2][X_1,X_2]),
\ee

\noindent 
where $X_1$ and $X_2$ are two $N \times N$ hermitean matrices, and $P_1$ and $P_2$ their conjugate momenta, respectively. 


One can think of (\ref{Ham12}) as associated with two of the six Higgs scalars of the bosonic sector of 
$\cal{N}=$ $4$ SYM, in the leading Kaluza Klein compactification on $S^3 \times R$. The harmonic potential results 
from the coupling to the curvature of the manifold. It should be borne in mind that we do not require supersymmetry.
Alternatively, compactification of $QCD_{2+1}$ on a sphere results in a similar hamiltonian.


We will follow the approach first suggested in \cite{Donos:2005vm} of treating one of matrices, $X_1$, in coordinate space and exactly (in the large $N$ limit), and the other, $X_2$, in a creation/annihilation basis. 
Letting

\be\label{Ide}
                             X_2 \equiv \frac{1}{\sqrt{2w}}( A_2 + A_2^{\dagger})  \quad P_2 = -i \sqrt{\frac{w}{2}}( A_2 - A_2^{\dagger}) 
\ee

\noindent
the hamiltonian (\ref{Ham12}) takes the form

\bea\label{HamMB}
\hat{H} &=& 
\frac{1}{2} \tr (P_1^2 ) + \frac{w^2}{2}  \tr (X_1^2) + w  \tr (A_2^{\dagger} A_2) + N^2 \frac{w}{2} \nonumber \\
&-& \frac {g_{YM}^2}{2w} \tr ( 2 [X_1,A_2^{\dagger}][X_1,A_2] + [X_1,A_2]^2  + [X_1,A_2^{\dagger}]^2 )  \\ 
&+& \frac {g_{YM}^2N}{w} \tr (X_1^2) -   \frac {g_{YM}^2}{w} (\tr (X_1))^2 \nonumber             \eea

\noindent
As the interaction is quadratic in the oscillators, one can perform a Bogoliubov transformation

\be\label{OscDef}
   (V^{\dagger} {A_2} V)_{ij} = \cosh (\phi_{ij}) B_{ij} - \sinh (\phi_{ij}) B^{\dagger}_{ij}  
\ee

\noindent
with

\be\label{OscAngle}
 \tanh(2\phi_{ij}) = \frac{\frac {g_{YM}^2}{w}(\lambda_i-\lambda_j)^2}{w + \frac{g_{YM}^2}{w}(\lambda_i-\lambda_j)^2},
\ee

\noindent
where the $\lambda_i$'s are the eigenvalues of the matrix $X_1$ and $V$ is the unitary matrix that diagonalizes $X_1$.
Then (\ref{HamMB}) takes the form

\be\label{Htot}
\hat{H} = \frac{1}{2} \tr (P_1^2 ) + \frac{w^2}{2}  \tr (X_1^2) +
\sum_{i,j=1}^N \sqrt{w^2 + 2 {g_{YM}^2}(\lambda_i-\lambda_j)^2} 
\quad({B}^{\dagger}_{ij}{B}_{ji} + \frac{1}{2}).
\ee


\noindent
We are interested in the leading large $N$ configuration of the system. In the $B$,$B^{\dagger}$ sector of the theory, states with non zero $B$ quanta lead to excited states and spectra, which contributions which are subleading in $\frac{1}{N}$. The only contribution to the large $N$ ground state configuration comes from the zero point energies of the $B$,$B^{\dagger}$ oscillators, and we are therefore led to the Hamiltonian:

\be\label{Hzero}
\hat{H_0} = \frac{1}{2} \tr (P_1^2 ) + \frac{w^2}{2}  \tr (X_1^2) +
\frac{1}{2} \sum_{i,j=1}^N \sqrt{w^2 + 2 {g_{YM}^2}(\lambda_i-\lambda_j)^2} 
\ee

\section{A self consistent non linear integral equation for the large $N$ background}

The Hamiltonian (\ref{Hzero}) is describes the dynamics of a single hermitean matrix, and the large $N$ background can be described in terms of the density of eigenvalues,

$$
\phi(x) = \sum_i \delta (x-\lambda_i),
$$

\noindent
as the minimum of the cubic collective field effective potential \cite{Jevicki:1979mb}

\bea\label{CollPot}
V_{eff} &=& {\frac{\pi^2}{6}} \int dx \phi^3(x) +  \frac{w^2}{2} \int dx \phi(x) x^2  - \mu (\int dx \phi(x) - N) \nonumber \\
&+& \frac{1}{2} \int dx \int dy \sqrt{w^2 + 2 {g_{YM}^2}(x-y)^2} \quad \phi(x) \phi(y) ,
\eea

\noindent
where the Lagrange multiplier $\mu$ enforces the contraint $\int dx \phi(x) = N$.
To exhibit explicitly the $N$ dependence, we rescale

\be \label{Rescaling}
x  \to \sqrt{N} \quad  \phi(x) \to \sqrt{N} \phi(x)  \quad  \mu \to N \mu\ 
\ee

\noindent
and obtain 

\bea\label{CollPotResc}
V_{eff} &=& N^2 \Big[ {\frac{\pi^2}{6}} \int dx \phi^3(x) +  \frac{w^2}{2} \int dx \phi(x) x^2 - \mu (\int dx \phi(x) - 1) \nonumber \\
&+& \frac{1}{2} \int dx \int dy \sqrt{w^2 + 2 \lambda (x-y)^2} \phi(x) \phi(y) \Big]
\eea

\noindent
where $\lambda=g_{YM}^2 N$ is the usual 't Hooft's coupling. 

As $N \to \infty$, the large N background configuration minimizes (\ref{CollPotResc}) and it satisfies:

\be\label{SelfCons}
\pi^2 \phi_0^2(x) = 2 \mu - w^2 x^2 - 2 \int dy \sqrt{w^2 + 2 \lambda (x-y)^2} \phi_0(y)
\ee

\section{Perturbative expansion}

\noindent
When $\lambda=0$, (\ref{SelfCons}) reduces to the well known Wigner distribution:

\be\label{po}
\pi \phi_0(x) = \sqrt{2\mu - 2w -w^2 x^2} = \sqrt {2 w - w^2 x^2}, \quad   |x| \le x_0 = \sqrt {\frac{2}{w}} 
\ee

\noindent
with the identification $\mu=2w$ being enforced by the constraint. Given that $\frac{2}{\pi} \int_0^{x_0} dx x^2 \sqrt {2 w - w^2 x^2} = \frac{1}{2w}$ it is straightforward to obtain   

\bea
E_0 &=& N^2 \Big[ {\frac{\pi^2}{6}} \int dx \phi_0^3(x) +  \frac{w^2}{2} \int dx \phi_0 (x) x^2 + \frac{w}{2} (\int dx \phi_0(x))^2 \Big] \nonumber \\
&=& N^2 \Big[ \frac{w}{4} + \frac{w}{4} + \frac{w}{2} \Big] = N^2 w 
\eea

\noindent
as it should, the ground state energy being simply the sum of the zero point energies of two ($N^2$) free oscillators.

To next order, we assume that the background remains even ($\int dx x \phi_0=0 $) and obtain from (\ref{SelfCons}) 

\be\label{FirstOrd}
\pi^2 \phi_0^2(x) = 2 \mu - 2 w - \frac{2 \lambda}{w} \omega_2 - (w^2 + \frac{2 \lambda}{w}) x^2, \quad \omega_2 = \int x^2 \phi_0(x)
\ee

\noindent
Therefore, to this order, the background remains of the Wigner type\footnote{$\mu$ is suitably adjusted.}:

\be\label{pio}
\pi \phi_0(x) = \sqrt {2 \bar w - {\bar w}^2 x^2}, \quad   |x| \le x_0 = \sqrt {\frac{2}{\bar w}} \quad {\bar w}^2 =w^2 + \frac{2 \lambda}{w}
\ee

\noindent
It follows that

$$
E_0 = N^2 \Big[ \frac{\bar w}{4} + \frac{\bar w}{4} + \frac{w}{2} \Big] = N^2 \Big[ w + \frac{\lambda}{2 w^2} + ... \Big] .
$$

\noindent
As an example of an equal time correlator, one straightforwardly obtains:

\be\label{CorOne}
<Tr X_1^2 > = N^2 \int dx x^2 \phi_0 = N^2 \frac{1}{2\bar{w}} = N^2 \Big[ \frac{1}{2w} - \frac{\lambda}{2 w^4} + ... \Big] 
\ee

\noindent
These results are in agreement with perturbation theory. 

Planar expectation values of correlators involving the $X_2$ coordinate can also be calculated. This requires expressing $X_2$ in terms of the $B$ oscillators 
using  (\ref{OscDef}) and (\ref{OscAngle}), and then reducing the correlator to a correlator involving the $X_1$ matrix only, 
with the use of the $B$ oscillator commutation relations. Because the $B, B^{\dagger}$ sector is quadratic in the oscillators, 
this can lead to highly nontrivial non-perturbative results. For instance,

\bea\label{CorTwo}
<Tr X_2^2 > &=& \frac{1}{2w} \sum_{ij} \cosh (2\phi_{ij})  - \sinh (2\phi_{ij}) =
\frac{1}{2} \sum_{ij} \frac{1}{\sqrt{w^2 + 2 g_{YM}^2 (\lambda_i-\lambda_j)^2}} \nonumber \\
&=&\frac{N^2}{2}\int dx \int dy \frac{\phi_0(x) \phi_0(y)}{\sqrt{w^2 + 2 \lambda (x-y)^2}}
= \frac{N^2}{2w} \Big( 1 - \frac{\lambda}{ w^3} + ... \Big)  
\eea

This equation shows how, to order $\lambda$, the $X_1 \leftrightarrow X_2$ symmetry of planar correlators is satisfied, despite the asymmetric treatment 
of the two coordinates in the approach followed in this communication. If we require, as we must, that this $X_1 \leftrightarrow X_2$ symmetry is exact, 
then (\ref{CorOne}) and (\ref{CorTwo}) establish a highly non-trivial property of the planar background.

Of great physical interest is to obtain the states which correspond to the spectrum of the system. This can be done in principle using the framework
developed in \cite{Donos:2005vm}, \cite{Rodrigues:2005ec} and \cite{Cook:2007et}, but is beyond the scope of this communication.

\section {Strong coupling solution}

We consider now (\ref{SelfCons}) and (\ref{CollPotResc}) as $\lambda \to \infty$ :

\be\label{SelfLarge}
\pi^2 \phi_0^2(x) = 2 \mu - 2 \sqrt{2\lambda}\int dy |x-y| \phi_0(y)
\ee

\be\label{CollPotLarge}
E_0 = N^2 \Big[ {\frac{\pi^2}{6}} \int dx \phi_0^3(x) + \frac{\sqrt{2\lambda}}{2} \int dx \int dy |x-y| \phi_0(x) \phi_0(y) \Big]
\ee

\noindent
We find it useful to introduce

\be\label{fdef}
  f(x) = \sqrt{2\lambda}\int dy |x-y| \phi_0(y),  \quad  \pi^2 \phi_0^2(x) = 2 (\mu - f(x)) 
\ee

\noindent
which satisfies 

\be\label{integeq}
f(x) = \frac{\sqrt{2\lambda}}{\pi}\int dy |x-y| \sqrt{ 2 (\mu - f(y))}.
\ee

As it was the case in perturbation theory, we assume that $\phi_0(x)$ remains an even, single cut function defined in the interval 
$[-x_0,x_0]$. To show that this is a consistent ansatz, we note that then:

\be\label{fintrep}
f(x) = \sqrt{2\lambda} \left( |x| \int_{-|x|}^{|x|} \phi_0(y) dy + 2 \int_{|x|}^{x_0} \phi_0(y) y dy \right).
\ee

Hence $f(x)$ is also even, establishing the consistency of the ansatz.
 
\noindent
Using

$$
           \partial_x^2 |x-y| = 2 \delta (x-y),
$$

\noindent
equation (\ref{integeq}) becomes\footnote{$\phi_0^2$ satisfies a very similar equation.}

\be\label{nonlineq}
  \partial_x^2 f(x) = \frac{4\sqrt{\lambda}}{\pi} \sqrt{\mu - f(x)}                              
\ee

This can be integrated in the usual way, to yield:

\be\label{enerlike}
  \frac{1}{2} (\partial_x f)^2 + \frac{8\sqrt{\lambda}}{3 \pi} (\mu - f(x))^{\frac{3}{2}} = e                              
\ee
 
The "energy" constant is fixed by the condition $\partial_x f(0)=0$. Denoting $f(0)\equiv f_0$ one obtains

\be\label{derone}
\frac{df}{dx} = 
\frac{4{\lambda}^{\frac{1}{4}}}{\sqrt{3\pi}} \sqrt{(\mu - f_0)^{\frac{3}{2}}- (\mu - f(x))^{\frac{3}{2}}}
\ee

The normalization condition 

$$
     1 = \int_{-x_0}^{x_0} dx \phi_0(x) = 2 \int_{0}^{x_0} dx \phi_0(x) = 2 \int_{f_0}^{\mu} df \frac{\phi_0(f)}{\frac{df}{dx}}
$$

\noindent
fixes

$$
                      (\mu - f_0)^{\frac{3}{2}} = (\frac{3\pi}{8})\lambda^{\frac{1}{2}} ,
$$

\noindent
and hence (\ref{derone}) takes the form:
  
\be\label{dersec}
\frac{df}{dx} = 
\sqrt{2{\lambda}}       \sqrt{1  - \Big( \frac{\mu - f(x)}{\mu - f_0} \Big)^{\frac{3}{2}}} .
\ee

We will not need to invert (\ref{dersec}) and obtain $f(x)$ explicitly, as all results presented in this communication will be expressed in terms of known definite integrals.

Of particular interest is the large $N$ ground state energy. From (\ref{CollPotLarge}) and (\ref{fdef}) this can be written as 

\be\label{CollPotCal}
E_0 = N^2 \Big[ {\frac{\pi^2}{6}} \int dx \phi_0^3(x) + \frac{1}{2} \int dx f(x) \phi_0(x)  \Big] =
N^2 \Big[ \frac{\mu}{2} - {\frac{\pi^2}{12}} \int dx \phi_0^3(x)\Big]
\ee

One needs to know $\mu$, or $f_0$, independently. From (\ref{fintrep}), one obtains

$$
          f_0 = \sqrt{2\lambda} x_0 - (\mu - f_0) \qquad \mu = \sqrt{2\lambda} x_0 .
$$  

From (\ref{dersec}) one obtains

$$
\sqrt{2\lambda} x_0 = (\mu - f_0)  \int_0^1 \frac{dz}{\sqrt{1 - (1-z)^{\frac{3}{2}}}} = 2 (\mu - f_0)  \int_0^1 \frac{t dt}{\sqrt{1 - t^3}} .
$$

Also

$$
{\frac{\pi^2}{12}} \int dx \phi_0^3(x) = \frac{1}{6} (\mu - f_0)  \int_0^1 dz \sqrt{1 - (1-z)^{\frac{3}{2}}} = 
\frac{1}{3} (\mu - f_0)  \int_0^1 t dt \sqrt{1 - t^3} .
$$

These integrals are tabulated \cite{Gradshteyn}, and are finite. Therefore 

\be\label{StrEn}
E_0 = N^2 \Big[ \frac{6}{7} ~\big( \frac{3\pi}{8} \big)^{\frac{2}{3}}  \int_0^1 \frac{t dt}{\sqrt{1 - t^3}}  ~\lambda^{\frac{1}{3}} \Big]
= N^2 \Big[ \frac{9}{14} \Big( \frac{\sqrt{3}}{4\pi} \Big)^{\frac{1}{3}} \Big( \Gamma \big( \frac{2}{3} \big) \Big)^{3} ~\lambda^{\frac{1}{3}} \Big]
\ee

Similarly to the weak coupling case, we consider the correlator

$$
<Tr X_1^2 > = N^2 \int dx x^2 \phi_0= 2 N^2 \int_{f_0}^{\mu} x^2 (f) \frac{\phi_0(f)}{\frac{df}{dx}} df
$$

By a sequence of integrations by parts, we obtain

\bea\label{StrCor}
<Tr X_1^2 > &=& N^2 \Big[ - \frac{\mu^2}{2\lambda} + \frac{2}{\sqrt{2\lambda}} \int_{f_0}^{\mu} \frac{f}{\frac{df}{dx}} df \Big]  \\
&=& 2 ~N^2  ~\big( \frac{3\pi}{8} \big)^{\frac{4}{3}} ~\lambda^{-{\frac{1}{3}}}
\Big[ \Big( \int_0^1 \frac{t dt}{\sqrt{1 - t^3}} \Big)^2 - \frac{2}{5} \int_0^1 \frac{dt}{\sqrt{1 - t^3}} \Big] \nonumber \\
&=&\frac{N^2}{\pi 2^{\frac{1}{3}} \sqrt{3}} \big( \frac{3\pi}{8} \big)^{\frac{4}{3}} ~\lambda^{-{\frac{1}{3}}}
\Big[ \frac{3\sqrt{3}}{\pi}     \Big( \Gamma \big( \frac{2}{3} \big) \Big)^{6} - \frac{2}{5} \Big( \Gamma \big( \frac{1}{3} \big) \Big)^{3} \Big] \nonumber
\eea

\section{Interacting massless matrices}

The results of the previous section could have also been obtained by taking the limit $w \to 0$ of the planar hamiltonian (\ref{Hzero}). It is therefore relevant
to discuss the relevance of these results for the system

\be\label{HamFree}
\hat{H}
\equiv
\frac{1}{2} Tr (P_1^2 ) + \frac{1}{2} Tr (P_2^2 ) - g_{YM}^2 \tr ([X_1,X_2][X_1,X_2]),
\ee

\noindent
which results the from the dimensional reduction of massless Higgs or non abelian vector potentials. The hamiltonian (\ref{HamFree}) has a single dimensionful
parameter, $\lambda = g_{YM}^2 N$, and therefore all observable quantities should depend only on well defined powers of $\lambda$. However, as is well known, 
perturbation theory is plagued
with infrared divergences. In this context, $w$ can thought of as a standard  ``mass" regulator.

The results of the previous section are therefore remarkable, as they are finite and free of any infrared divergences, and depend only on the appropriate 
power of $\lambda$ which is expected from dimension considerations. 

It has already been pointed out the expression (\ref{Hzero}) has a smooth $w \to 0$ limit. 
The only place where this limit is potentially ill defined is in the transformation (\ref{OscDef}), where where $w$ has to be kept finite, if small, 
to ensure that the tranformation is canonical. However, once the Bogoliubov tranformation is implemented and the Hamiltonian (\ref{Ham12}) is recast in the
form  (\ref{Htot}), its $w \to 0$ limit should provide a correct description of the system (\ref{HamFree}).       

We are confident that indeed the results of the previous section, which are finite and free of any infrared divergences, are the planar energy and correlator
of $(\ref{HamFree})$. This provides another explicit confirmation of the expectation that strong coupling dynamics, appropriately resummed through
the large N limit, is free of infrared divergences.

\section{Summary and discussion}

In this communication, we obtained the large $N$ planar background of two hermitean matrices, in an harmonic potential, interacting through a Yang-Mills potential, 
in terms of the density of eigenvalues of one of the matrices. This background is shown to satisfy a self-consistent $g_{YM}$ dependent integral equation. 
This integral equation has both a weak and a strong coupling expansion. The weak coupling expansion of the backround is described and shown to be in agreement with perturbation theory. 
For strong coupling, we show that the background satisfies a non-linear differential equation, with a solution which is described. 
The planar ground state energy and examples
of correlators are obtained and shown to be finite. We argue that this is the full planar solution of the hamiltonian of two ``massless" matrices
(i.e., without the harmonic potential, or in the zero curvature limit) with a Yang-Mills interaction.

It may be tempting to associate the background obtained in this communication with a non-supersymetric $g_{YM}$ deformation of the ``droplet" description of
$1/2$ BPS states \cite{Corley:2001zk}, \cite{Berenstein:2004kk}, \cite{Lin:2004nb}. It should be remembered, though, that the matrix description of the geometry 
is a phase space description. The hermitean matrix associated with the droplet corresponds to the holomorphic restriction of a complex matrix \cite{Donos:2005vm}, 
unlike the choice made in this communication\footnote{As an example of how the kinetic energy operator of the hamiltonian is changed in this case, see \cite{Rodrigues:2006gt}.} 

On the other hand, as pointed out at the end of Section $4$, by exploiting the $X_1 \leftrightarrow X_2$ symmetry of the system, we can write from (\ref{CorTwo})
and as $\lambda \to \infty$,

\be\label{Back}
<Tr X_1^2 > = <Tr X_2^2 > = \frac{1}{2} \sum_{ij} \frac{1}{\sqrt{2 g_{YM}^2} ~|\lambda_i-\lambda_j|} .
\ee

\noindent
With a physical interpretation of the eigenvalues as coordinates of a system of $D0$'s, the feature of (\ref{StrCor}) that its ``size" becomes small but finite
for large $\lambda$ can be interpreted as resulting from the ``back-reaction" exhibited on the right hand side of (\ref{Back}). 

Clearly, it would be very interesting to examine this question further. However, at the very least, the background
identified in this article is a $g_{YM}$ deformation of the Wigner distribution associated with harmonic potentials, with a well defined strong coupling limit. 

\section{Acknowledgements}
One of us (J.P.R.) would like to thank the High Energy Theory Group of Brown University for their hospitality during several research visits over the last two years,
which allowed for concentrated effort leading to some of the ideas presented in this communication. We thank Antal Jevicki and Robert de Mello Koch for 
reading the manuscript, comments and discussion.

\end{document}